\documentstyle[11pt,epsfig]{article}
\title{\bf  Modified gravity inspired DGP brane cosmology}
\author{K. Atazadeh and H. R.
Sepangi\thanks{email: hr-sepangi@sbu.ac.ir}
\\ {\small Department of Physics, Shahid Beheshti University, Evin,
Tehran 19839, Iran}}
\begin{document}
\maketitle

\begin{abstract}
We consider a DGP brane scenario where a scalar field is present
on the brane through the introduction of a scalar potential,
itself motivated by the notion of modified gravity. This theory
predicts that the mass appearing in the gravitational potential is
modified by the addition of the mass of the scalar field. The
cosmological implications that such a scenario entails are
examined and shown to be consistent with a universe expanding with
power-law acceleration.
\end{abstract}
\vspace{2cm}

\section{Introduction}
The idea that extra dimensions can be probed by gravitons and
eventually non-standard matter has been the dominant trend in the
recent past. These models usually yield the correct Newtonian
$(1/r)$ potential at large distances because the gravitational
field is quenched on sub-millimeter transverse scales. This
quenching appears either due to finite extension of the transverse
dimensions \cite{1,2} or due to sub-millimeter transverse
curvature scales induced by negative cosmological constants
\cite{3,4,5,6,7,8}. A feature common to these type of models is
that they predict deviations from the usual $4D$ gravity at short
distances. The model proposed by Dvali, Gabadadze and Porrati
(DGP) \cite{10,11} is very different in that it predicts
deviations from the standard $4D$ gravity over large distances.
The transition between four  and higher-dimensional gravitational
potentials in the DGP model arises because of the presence of both
the brane and bulk Einstein terms in the action. An interesting
observation was made in \cite{13,14} where it was shown that the
DGP model allows for an embedding of the standard Friedmann
cosmology in the sense that the cosmological evolution of the
background metric on the brane can entirely be described by the
standard Friedmann equation plus energy conservation on the brane.
This was later generalized to arbitrary number of transverse
dimensions in \cite{15}. For a comprehensive review of the
phenomenology of DGP cosmology, the reader is referred to
\cite{lue}.

An interesting observation made a few years ago is that the
expansion of our universe is currently undergoing a period of
acceleration which is directly measured from the light-curves of
several hundred type Ia supernovae \cite{16,17} and independently
from observations of the cosmic microwave background (CMB) by the
WMAP satellite \cite{18} and other CMB experiments \cite{19}.
However, the mechanism responsible for this acceleration is not
well understood and many authors introduce a mysterious cosmic
fluid, the so called dark energy, to explain this effect
\cite{20}. Recently, it has been shown that such an accelerated
expansion could be the result of a modification to the
Einstein-Hilbert action \cite{21}. One such modification has been
proposed in \cite{22} where a term of the form $R^{-1}$ was added
to the usual Einstein-Hilbert action. It was then shown that this
term could give rise to accelerating solutions of the field
equations without dark energy.

In this paper, we focus attention on the DGP brane model and
introduce a scalar field on the brane. The potential describing
such a scalar field is taken to be that appearing in modified
theories of gravity when the term $R^{-1}$ is added to the usual
Einstein-Hilbert action. This model predicts that for such a
potential, the mass density should be modified by the addition of
the mass density of the corresponding scalar field on the brane.
We obtain the evolution of the metric on the spacetime by solving
the field equations in the limit of small curvature, predicting a
power-law acceleration on the brane. The components of the metric
in Gaussian normal coordinates are also calculated and presented.
\section{DGP model with a brane scalar field}
We start by writing the action for the DGP model with a scalar
field on the brane part of the action
\begin{equation}\label{eq1}
{\cal S}=\frac{m^{3}_{4}}{2}\int d^{5}x\sqrt{-g}{\cal R}+\int
d^{4}x\sqrt{-q}\left[\frac{m^{2}_{3}}{2}R-
\frac{1}{2}(\nabla\Phi)^{2}-V(\Phi)\right]+{\cal
S}_{m}\left[q_{\mu\nu},\psi_{m}\right].
\end{equation}
where the first term in (\ref{eq1}) corresponds to the
Einstein-Hilbert action in $5D$ for the 5-dimensional bulk metric
$g_{AB}$, with the Ricci scalar denoted by ${\cal R}$. Similarly,
the second term is the Einstein-Hilbert action for a scalar field
$\Phi$ corresponding to the induced metric $q_{\mu\nu}$ on the
brane, where $R$ is the relevant scalar curvature and $m_{3}$ and
$m_{4}$ are reduced Planck masses in four and five dimensions
respectively and ${\cal S}_{m}$ is the matter action on the brane
with matter field $\psi_{m}$. The induced metric $q_{\mu\nu}$ is
defined as usual from the bulk metric $g_{AB}$ by
\begin{equation}\label{eq2}
q_{\mu\nu}=\delta^{A}_{\mu}\delta^{B}_{\nu}g_{AB}.
\end{equation}
It would now be possible to write the field equations resulting from
this action, yielding, in $d-1$ spatial dimensions
\begin{equation}\label{eq3}
m^{3}_{4}\left({\cal R}_{AB}-\frac{1}{2}g_{AB}{\cal
R}\right)+m^{2}_{3}\delta^{\mu}_{A}\delta^{\nu}_{B}\left(R^{(d-1)}_{\mu\nu}-
\frac{1}{2}q_{\mu\nu}R^{(d-1)}\right)\delta(y)
=\delta^{\mu}_{A}\delta^{\nu}_{B}\left(T_{\mu\nu}+{\cal
T}_{\mu\nu}\right)\delta(y),
\end{equation}
where $T_{\mu\nu}$ is the energy-momentum tensor in the matter
frame, and
\begin{equation}\label{eq4}
{\cal
T}_{\mu\nu}=\partial_{\mu}\Phi\partial_{\nu}\Phi-\frac{1}{2}q_{\mu\nu}q^{\alpha\beta}
\partial_{\alpha}\Phi\partial_{\beta}\Phi
-V(\Phi)q_{\mu\nu},
\end{equation}
with the equation of motion for $\Phi$ becoming
\begin{equation}\label{eq5}
\nabla_{\mu}\nabla^{\mu}\Phi=\frac{dV(\Phi)}{d\Phi}.
\end{equation}
Note that $\Phi$ lives on the brane. The corresponding junction
conditions, relating the extrinsic curvature to the
energy-momentum tensor, become \cite{13,14}
\begin{eqnarray}\label{eq6}
\mbox{lim}_{\epsilon\rightarrow+0}[K_{\mu\nu}]^{y=+\epsilon}_{y=-\epsilon}
&=&\left.\frac{1}{m^{3}_{4}}\left(T_{\mu\nu}+{\cal
T}_{\mu\nu}-\frac{1}{d-1}q_{\mu\nu}q^{\alpha\beta}\left(T_{\alpha\beta}+{\cal
T}_{\alpha\beta}\right)\right)\right|_{y=0}\nonumber\\
&-&\left.\frac{m^{2}_{3}}{m^{3}_{4}}\left(R^{(d-1)}_{\mu\nu}-
\frac{1}{2(d-1)}q_{\mu\nu}q^{\alpha\beta}R^{(d-1)}_{\alpha\beta}\right)\right|_{y=0}.
\end{eqnarray}
In order to get a qualitative picture of how gravity works for the
DGP braneworld, let us take small metric fluctuations around flat,
empty space and look at gravitational perturbations, $h_{AB}$,
that is
\begin{equation}\label{eq7}
g_{AB}=\eta_{AB}+h_{AB},
\end{equation}
where $\eta_{AB}$ is the five-dimensional Minkowski metric. Choosing
the harmonic gauge in the bulk
\begin{equation}\label{eq8}
\partial^{A}h_{AB}=\frac{1}{2}\partial_{B}h^{A}_{A},
\end{equation}
the $\mu5$-components of this gauge condition lead to
$h_{\mu5}$=0, so that the surviving components are $h_{\mu\nu}$
and $h_{55}$. The latter component is solved by the following
equation
\begin{equation}\label{eq9}
\Box^{(5)}h^{5}_{5}=\Box^{(5)}h^{\mu}_{\mu},
\end{equation}
where $\Box^{(5)}$ is the five-dimensional d'Alembertian. The
$\mu\nu$-component of the field equations (\ref{eq3}) become,
after a little manipulation \cite{11}
\begin{equation}\label{eq10}
m^{3}_{4}\Box^{(5)}h_{\mu\nu}+m^{2}_{3}\left(\Box^{(4)}h_{\mu\nu}-\partial_{\mu}\partial_{\nu}h^{5}_{5}\right)\delta(y)=
-2\delta(y)\left[T_{\mu\nu}+{\cal
T}_{\mu\nu}-\frac{1}{d-1}\eta_{\mu\nu}\eta^{\alpha\beta}\left(T_{\alpha\beta}+{\cal
T}_{\alpha\beta}\right)\right],
\end{equation}
where $\Box^{(4)}$ is the four-dimensional (brane) d'Alembertian,
and we take the brane to be located at $y=0$. This yields the
equation for the gravitational potential of mass densities
$\rho(\vec{r})=M\delta(\vec{r})$ and
$\rho_{\Phi}(\vec{r})=M_{\Phi}\delta(\vec{r})$ on the brane
\begin{equation}\label{eq11}
m^{3}_{4}\left(\Box^{(4)}+\partial^{2}_{y}\right)U(\vec{r},y)+m^{2}_{3}\delta(y)\Box^{(4)}U(\vec{r},y)=\frac{2}{3}
(M+M_{\Phi})\delta(\vec{r})\delta(y).
\end{equation}
Therefore, in equation (\ref{eq11}), the mass is modified by the
addition of the mass of the scalar field. The resulting
gravitational potential for $r\ll\ell _{DGP}$ is given by
\cite{10,13}
\begin{equation}\label{eq13}
U(\vec{r})=-\frac{(M+M_{\Phi})}{6\pi
m^{2}_{3}r}\left[1+\left(\gamma-\frac{2}{\pi}\right)\frac{r}{\ell_{DGP}}+
\frac{r}{\ell_{DGP}}\ln\left(\frac{r}{\ell_{DGP}}\right)+{\cal
O}\left(\frac{r}{\ell_{DGP}}\right)^{2}\right],
\end{equation}
and
\begin{equation}\label{eq14}
U(\vec{r})=-\frac{(M+M_{\Phi})}{6\pi^{2}
m^{3}_{4}r^{2}}\left[1-2\left(\frac{r}{\ell_{DGP}}\right)^{-2}+{\cal
O}\left(\frac{r}{\ell_{DGP}}\right)^{-4}\right],
\end{equation}
for $r\gg\ell_{DGP}$, where $\gamma=0.577$ is the Euler constant
and $\ell _{DGP}=\frac{m_{3}^{2}}{2m_{4}^{3}}$ is the transition
scale between the four and five-dimensional behavior of
gravitational potential that the DGP scenario predicts.
\section{DGP cosmology with a brane scalar field}
Although the DGP model predicts deviations to gravity at large
distances, it could account for the standard cosmological
equations of motion at any distance scale on the brane. It is
therefore appropriate to start by writing  the from of the line
element in brane gravity, that is
\begin{equation}\label{eq15}
ds^{2}=q_{\mu\nu}dx^{\mu}dx^{\nu}+b^{2}(y,t)dy^{2}=-n^{2}(y,t)dt^{2}+
a^{2}(y,t)\gamma_{ij}dx^{i}dx^{j}+b^{2}(y,t)dy^{2},
\end{equation}
where $\gamma_{ij}$ is a maximally symmetric 3-dimensional metric
where $k=-1,0,1$ parameterizes the spatial curvature. Building on
the results of \cite{26}, the cosmological evolution equations of
a 3-brane in a 5-dimensional bulk resulting from equations
(\ref{eq3}) and (\ref{eq6}) were presented in the first two
references in \cite{21}. Here, we will follow \cite{13,14} and
only give the results relevant to the present work for a brane of
dimension $\nu+1$ and the scalar field $\Phi$. A detailed
discussion on the derivation of these results can be found in the
said references. Adopting the Gaussian normal system gauge
\begin{equation}\label{eq16}
b^{2}(y,t)=1,
\end{equation}
the field equations on the brane for metric (\ref{eq15}) and $d
=\nu+1$ spatial dimensions are
\begin{equation}\label{eq17}
G^{(\nu)}_{00}=\frac{1}{2}\nu(\nu-1)n^{2}\left(\frac{\dot{a}^{2}}{n^{2}a^{2}}+
\frac{k}{a^{2}}\right)=\frac{1}{m^{\nu-1}_{\nu}}{\cal T}_{00},
\end{equation}
\begin{equation}\label{eq18}
G^{(\nu)}_{ij}=(\nu-1)\left(\frac{\dot{n}\dot{a}}{n^{3}a}-
\frac{\ddot{a}}{n^{2}a}\right)q_{ij}-\frac{1}{2}(\nu-1)(\nu-2)
n^{2}\left(\frac{\dot{a}^{2}}{n^{2}a^{2}}+
\frac{k}{a^{2}}\right)q_{ij}=\frac{1}{m^{\nu-1}_{\nu}}{\cal T}_{ij},
\end{equation}
\begin{equation}\label{eq19}
\nabla_{\mu}\nabla^{\mu}\Phi=\frac{dV(\Phi)}{d\Phi}.
\end{equation}
The junction conditions (\ref{eq6}) for an ideal fluid on the
brane, given by
\begin{equation}\label{eq24}
T_{\mu\nu}=(\rho+p)u_{\mu}u_{\nu}+pq_{\mu\nu},
\end{equation}
together with energy conservation  resulting from the vanishing of
\begin{eqnarray}
G_{05}=\nu\left(\frac{n'}{n}\frac{\dot{a}}{a}-\frac{\dot{a}'}{a}\right)=0,\label{neq1}
\end{eqnarray}
in the bulk, leads to
\begin{equation}\label{eq31}
\left.\left(\dot{\rho}+\dot{\rho}_{\Phi}\right)a\right|_{y=0}=
-\nu\left.\left(\rho+\rho_{\Phi}+p+p_{\Phi}\right)\dot{a}\right|_{y=0},
\end{equation}
where
\begin{equation}\label{eq27}
\rho_{\Phi}=\left[\frac{1}{2}\dot{\Phi}^{2}+n^{2}V(\Phi)\right]_{y=0},
\end{equation}
\begin{equation}\label{eq28}
p_{\Phi}=a^{2}\left[\frac{1}{2n^{2}}\dot{\Phi}^{2}-V(\Phi)\right]_{y=0}.
\end{equation}
One may now proceed to obtain the cosmological equations by taking
the gauge
\begin{equation}\label{eq37}
n(0,t)=1,
\end{equation}
and performing the transformation
\begin{equation}\label{eq38}
t=\int^{t}n(0,\tau)d\tau,
\end{equation}
of the time coordinate. This gauge is convenient because it gives
the usual cosmological time on the brane. Consequently,  we find
that our basic dynamical variable is $a(y,t)$ with $n(y,t)$ given
by
\begin{equation}\label{eq39}
n(y,t)=\frac{\dot{a}(y,t)}{\dot{a}(0,t)}.
\end{equation}
The basic set of cosmological equations in the present setting for
a brane scalar field without a cosmological constant in the bulk
now become
\begin{eqnarray}\label{eq40}
\lim_{\epsilon\rightarrow+0}\left[\partial_{y}a\right]^{y=
+\epsilon}_{y=-\epsilon}(t)&=&\frac{m^{\nu-1}_{\nu}}{2m^{\nu}_{\nu+1}}(\nu-1)
\left[\frac{\dot{a}^{2}(0,t)}{a(0,t)}+
\left.\frac{k}{a(0,t)}\right]\right|_{y=0} \nonumber\\
&-&\left.\frac{(\rho+\rho_{\Phi})a(0,t)}{\nu
m^{\nu}_{\nu+1}}\right|_{y=0},
\end{eqnarray}
\begin{equation}\label{eq41}
I^{+}=\left.\left[\dot{a}^{2}(0,t)-a'^{2}(y,t)+k\right]a^{\nu-1}(y,t)\right|_{y>0},
\end{equation}
\begin{equation}\label{eq42}
I^{-}=\left.\left[\dot{a}^{2}(0,t)-a'^{2}(y,t)+k\right]a^{\nu-1}(y,t)\right|_{y<0},
\end{equation}
\begin{equation}\label{eq43}
\nabla_{\mu}\nabla^{\mu}\Phi=\frac{dV(\Phi)}{d\Phi},
\end{equation}
\begin{equation}\label{eq44}
n(y,t)=\frac{\dot{a}(y,t)}{\dot{a}(0,t)}.
\end{equation}

It is appropriate at this point to discuss the cosmology in the
DGP model by taking $I^{+}=I^{-}$. The cosmological equations in
this framework for a $(\nu-1)$-dimensional space are given by
\begin{equation}\label{eq45}
\frac{\dot{a}^{2}(0,t)+k}{a^{2}(0,t)}=
\frac{2(\rho+\rho_{\Phi})}{\nu(\nu-1)m^{\nu-1}_{\nu}},
\end{equation}
\begin{equation}\label{eq46}
\ddot{\Phi}+3\frac{\dot{a}(0,t)}{a(0,t)}\dot{\Phi}+\frac{dV(\Phi)}{d\Phi}
=0,
\end{equation}
\begin{equation}\label{eq47}
I=\left[\dot{a}^{2}(0,t)-a'^{2}(y,t)+k\right]a^{\nu-1}(y,t),
\end{equation}
\begin{equation}\label{eq48}
n(y,t)=\frac{\dot{a}(y,t)}{\dot{a}(0,t)}.
\end{equation}
Equations (\ref{eq47}) and (\ref{eq48}) may now be used, taking
$\nu=3$, to obtain the components of the metric
\begin{equation}\label{eq51}
a^{2}(y,t)=a^{2}(0,t)+
(\dot{a}^{2}(0,t)+k)y^{2}+2\sqrt{(\dot{a}^{2}(0,t)+k)
a^{2}(0,t)-I}y,
\end{equation}
\begin{equation}\label{eq52}
n(y,t)=\left[a(0,t)+ \ddot{a}(0,t)y^{2}+a(0,t)y
\frac{a(0,t)\ddot{a}(0,t)+
\dot{a}^{2}(0,t)+k}{\sqrt{(\dot{a}^{2}(0,t)+k)
a^{2}(0,t)-I}}\right]\frac{1}{a(y,t)}.
\end{equation}
The form of these equations becomes particularly simple for $I=
0$, that is
\begin{equation}\label{eq53}
a(y,t)=a(0,t)+\sqrt{\dot{a}^{2}(0,t)+k}y,
\end{equation}
\begin{equation}\label{eq54}
n(y,t)=1+\frac{\ddot{a}(0,t)}{\sqrt{\dot{a}^{2}(0,t)+k}}y.
\end{equation}

\section{$R^{-1}$ inspired DGP scenario}

To progress further, the form of the potential $V(\Phi)$ should be
specified. Motivated by theories of modified gravity where a
$R^{-1}$ term is present in the action and with an eye on the
effects that such a modification may have on the DGP brane
scenarios we take the potential as \cite{22}
\begin{equation}\label{eq55}
V(\Phi)\simeq\mu^{2}m^{2}_{3}\exp\left(-\sqrt{3/2}\Phi/m_{3}\right).
\end{equation}
This is the form of the potential one encounters when studying
theories of modified gravity with a Lagrangian of the form ${\cal
L}(R)=R-\frac{\mu^{4}}{R}$ and we shall concentrate in the limit
of small $R$ in the Einstein frame. Our aim is to obtain
explicitly the components of the metric on the brane, {\it i.e.}
equations (\ref{eq51}) and (\ref{eq52}) in the limit of small
curvature, and show that the results are consistent with an
accelerating universe. Note that one may obtain an effective mass
for the scalar field, $M^{eff}_{\Phi}$, from the second derivative
of the potential $V(\Phi)$  in the Einstein frame in the usual
$4D$ gravity with $\mu\sim H_{0}\simeq10^{-42}$ Gev and evaluating
$V''(\Phi)$ around $R\sim H_{0}^{2}$, noting that
$\Phi/m_{3}\sim1$. This leads to an effective mass squared of
order $\mu^{2}$, where $\mu$ is the cosmological constant in
$R^{-1}$ gravity \cite{22}. Hence, for models of modified gravity
with any function of the Ricci scalar, ${\cal L}(R)$, within the
framework of the discussion at hand, we must take into account the
contribution of the vacuum energy (cosmological constant) to the
gravitational potential.

To proceed, we consider the evolution of the scale factor with
time on the brane. From equations (\ref{eq45}) and (\ref{eq46})
for the spatially flat FRW metric and setting $\nu=3$, we can
write
\begin{equation}\label{eq56}
3H_{0}^{2}=\frac{1}{m^{2}_{3}}(\rho+\rho_{\Phi}),
\end{equation}
and
\begin{equation}\label{eq57}
\ddot{\Phi}+3H_{0}\dot{\Phi}+\frac{dV(\Phi)}{d\Phi}=0,
\end{equation}
where $H_{0}\equiv\frac{\dot{a}(0,t)}{a(0,t)}$ and for $n(0,t)=1$
(on the brane) we have
\begin{equation}\label{eq58}
\rho_{\Phi}=\frac{1}{2}\dot{\Phi}^2+V(\Phi).
\end{equation}
We must now solve the system of equations (\ref{eq56}) and
(\ref{eq57}) with $\rho=0$. Substituting potential (\ref{eq55})
into equations (\ref{eq56}) and (\ref{eq57}), we obtain the
evolution of the scale factor on the brane
\begin{equation}\label{eq59}
a(0,t)\propto t^{4/3},
\end{equation}
together with that of the scalar field
\begin{equation}\label{eq60}
\Phi\propto-\frac{4}{3}\ln t.
\end{equation}
As can be seen, equation (\ref{eq59}) predicts a power-law
acceleration on the brane. This result is consistent with the
observational results similar to quintessence with the equation of
state parameter $-1<w_{DE}<-\frac{1}{3}$, \cite{30}.

To continue, we find the evolution of $a(y,t)$ and $n(y,t)$
everywhere in spacetime. Thus, using equations (\ref{eq59}) and
substituting into equations (\ref{eq51}) and (\ref{eq52}), one
finds
\begin{equation}\label{eq61}
a^{2}(y,t)=C^{2}\left[t^{8/3}+\left(\frac{16}{9}\right)t^{2/3}y^{2}\right]+
2\sqrt{\left(\frac{16}{9}\right)C^{4}t^{10/3}-I}y.
\end{equation}
and
\begin{equation}\label{eq62}
n(y,t)=C\left[t^{4/3}+\left(\frac{4}{9}\right)y^{2}t^{-2/3}+
t^{4/3}y\frac{\left(\frac{20}{9}\right)C^{2}t^{4/3}}{\sqrt{\left(\frac{16}{9}\right)C^{4}t^{10/3}-I}}
\right]\frac{1}{a(y,t)}.
\end{equation}
In the particular case $I=0$, we obtain
\begin{equation}\label{eq63}
a(y,t)=C\left[t^{4/3}+\frac{4}{3}t^{1/3}y\right],
\end{equation}
and
\begin{equation}\label{eq64}
n(y,t)=C\left[1+\frac{y}{3t}\right],
\end{equation}
where $C$ is the proportionality constant. Note that for $y=0$,
equations (\ref{eq63}) and (\ref{eq64}) reduce to (\ref{eq59}) and
$n(0,t)=1$ respectively. There appears coordinate singularities on
the space-like hypercone $ y=\pm 3t$. This is presumably a
consequence of the fact that the orthogonal geodesics emerging
from the brane (which we used to set up our Gaussian normal
system, $b^{2}=1$) do not cover the full five-dimensional
spacetime. 
\section{Conclusions}
Brane models provide a somewhat exotic, yet interesting extension
of our parameter space for gravitational theories. In this work we
have considered the DGP  model with a brane scalar field. The
scalar field was motivated and inspired by a desire to study the
effects of modified gravity, represented by a term  like $R^{-1}$.
We have shown that this model predicts that the mass in the
gravitational potential is modified by the addition of the mass of
such a scalar field. The cosmological evolution of this model was
also studied by solving the relevant dynamical equations and the
components of the metric  was obtained in the limit of small
curvature. The evolution of the universe in such a scenario seems
to be consistent with the present observations, predicting an
accelerated expansion.

\end{document}